\documentstyle[aps,prb,epsf,floats]{revtex}

\begin{document}
\draft

\twocolumn[\hsize\textwidth\columnwidth\hsize\csname@twocolumnfalse%
\endcsname

\title{Excitations in antiferromagnetic cores of superconducting vortices}
\author{Henrik Bruus, Kasper Astrup Eriksen, 
Michael Hallundb\ae k and Per Hedeg\aa rd}
\address{\O rsted Laboratory, Niels Bohr Institute for APG, \\
Universitetsparken 5, DK-2100 Copenhagen \O\ Denmark}
\date{Submitted to Phys.\ Rev. B, July 10, 1998}
\maketitle
\begin{abstract}
We study excitations of the predicted antiferromagnetically ordered
vortex cores in the superconducting phase of the newly proposed $SO(5)$
model of strongly correlated electrons. Using experimental data from the
literature we show that the susceptibilities in the spin sector and
the charge sector are nearly equal, and likewise for the stiffnesses. 
In the case of strict equality $SO(5)$ symmetry is possible, and we
find that if present the vortices give rise to an enhanced neutron
scattering cross section near the so called $\pi$ resonance at
41~meV. In the case of broken $SO(5)$ symmetry two effects are
predicted. Bound excitations can exist in the vortex cores with
``high'' excitation energies slightly below 41~meV, and the massless
Goldstone modes corresponding to the antiferromagnetic ordering of the
core can acquire a mass and show up as core excitation with ``low''
excitation energies around 2~meV.

\end{abstract}
]

\section{Introduction}
\label{sec:intro}

Inspired by the discovery of a sharp antiferromagnetic resonance,
later denoted the $\pi$ resonance, in neutron scattering experiments
on the superconducting phase of YBa$_2$Cu$_3$O$_7$ at $(\pi,\pi)$ in
the reciprocal space \cite{mook,fong} a new idea was introduced a year
ago in the search for a theory of the high-$T_c$
superconductors. Within the framework of the $t$-$J$ limit of the
Hubbard model \cite{pwa,rice} it was shown theoretically already three
years ago \cite{demler} that the $\pi$ resonance could be
explained in terms of a new collective mode in the particle particle
channel of the model, and that this mode is intimately connected with
the symmetry of the superconducting gap. Pursuing the symmetry aspects
of the problem Zhang proposed a theory combining antiferromagnetism
and superconductivity by symmetry arguments. \cite{zhang} The
operators responsible for the $\pi$ resonance was identified with the
six generators of rotation between the antiferromagnetic state and the
superconducting state. Furthermore, it was proposed that the phase
diagram of the cuprates can be understood as a competition at low
temperatures between $d$-wave superconductivity and antiferromagnetism
of a system which at higher temperatures posses $SO(5)$ symmetry. The
group $SO(5)$ is sufficiently large to 
accommodate both the gauge group $U(1)$ ($=SO(2)$) which is broken in
the superconducting state, and the spin rotation group $SO(3)$ which
is broken in the antiferromagnetic state. In the simplest version the
transition between the two ordered states is controlled by one
parameter -- the chemical potential for holes in the otherwise
half-filled quadratic lattice of spin $1/2$ fermions. The idea has
generated a lot of discussion among
theorists\cite{baskaran,greiter,zhang1,laughlin}, and no consensus on
the matter has emerged. 

We shall not enter this discussion here. Rather we
will take the approach of assuming the model to be a fair description of
the high-$T_c$ materials and derive experimental consequences, which can
be tested in the laboratory. Arovas {\it et al.}\cite{arovas} have
pointed out that in the vortex cores of fluxoids in the superconducting
state the order parameter will escape into the antiferromagnetic
subspace, meaning that in these cores we have local antiferromagnetic
moments instead of a simple featureless normal metal core. This unique
prediction of the $SO(5)$ model should in principle be quite simple to
verify experimentally. However, preliminary measurements
\cite{andersen} where one looked for Bragg-scattering from these
moments belonging to the vortex cores did not produce any signal. This
is perhaps not so surprising, since each of the vortex cores will form
a one-dimensional magnet (along the $c$ axis) and at finite
temperatures such a system does not form long range order, and no
Bragg peak is to be expected.

In this paper we are going to pursue the idea that in each
copper-oxide plane small islands of antiferromagnetically ordered
spins exist associated with the vortices generated by an external
magnetic field. The direction of the spins in these islands will not
be very strongly correlated from layer to layer and from island to
island. In one island, however, there should exist excitations of the
spins, a kind of bound spin wave modes or size quantized
magnons. Using samples in the under-doped regime of the
superconducting phase, where the proximity of the antiferromagnetic
phase stabilizes the antiferromagnetic vortex cores \cite{arovas}, one
should be able to pick up these core excitations in inelastic neutron
scattering measurements. The modes can be classified according to the
approximate symmetry: Two zero-energy or ``low'' energy Goldstone
modes whose existence is guarantied by the exact spin rotation
symmetry, and two resonances or weakly bound ``high'' energy modes
related to the $\pi$ resonance arising form the approximate $SO(5)$
symmetry allowing for rotations between the $d$ wave superconducting
phase and the antiferromagnetic phase. Based on experimental data 
taken from the literature we discuss in Sec.~\ref{sec:SO5} the $SO(5)$
model and its parameters. The values of the susceptibilities and the
stiffnesses in the charge sector and the spin sector are found to be
nearly equal, a remarkable fact supporting the $SO(5)$ model. 
In Sec.~\ref{sec:coreexci} we set up the
calculation for excitations of the vortex core. In the isotropic case
we show analytically that the Goldstone modes of the vortex core
indeed have zero energy, and that the vortex indeed generates a
resonance reminiscent of the $\pi$ resonance at the bottom of
the continuum. In the anisotropic case the Goldstone modes remains
massless, but for certain anisotropies the $\pi$ resonance can be  
transformed into a bound state localized at the vortex.
Treating the external fields more accurately introduces
symmetry-breaking terms in the Hamiltonian and results in finite
masses to the Goldstone modes. This will be discussed in
Sec.~\ref{sec:Goldstone}. Concluding remarks are contained in
Sec.~\ref{sec:conclusions}. 

\section{The $SO(5)$ model and its parameters}
\label{sec:SO5}

In the $SO(5)$ model the relevant order parameter is a 
real vector $\bf{n}$ in a five dimensional superspin space with a length
which is fixed at low temperatures,

\begin{equation} \label{eq:n_vector}
{\bf n} = \{n_1,n_2,n_3,n_4,n_5\}, \qquad |{\bf n}|^2 = 1.
\end{equation}
This order parameter is related to the complex superconductor
order parameter, $\psi$, and the antiferromagnetic order order
parameter, ${\bf m}$, in each copper-oxide plane as follows:

\begin{equation} \label{eq:orderparameter}
\psi =  f e^{i\theta} = n_1 + i n_5, \qquad 
{\bf m} = (n_2,n_3,n_4).
\end{equation}
In Ref.~\onlinecite{zhang} Zhang argued how in terms of the five
dimensional superspin space one can construct an effective Lagrangian
$\cal L({\bf n})$ describing the low energy physics of the $t$-$J$
model. In the absence of external electromagnetic fields it takes the
form

\begin{eqnarray} \label{eq:L_of_n}
&& {\cal L}({\bf n}) = 
\nonumber \\ 
&& \sum_{a<b} \frac{\chi_{ab}}{2}  \left[
n_a\!\left(\partial_tn_b-
   \frac{2\mu}{\hbar} \{\delta_{b,1}n_5-\delta_{b,5}n_1 \}\right) -
\left( \rule{0mm}{1.5em} \!a \leftrightarrow b\! \right) \right]^2 
\nonumber \\ 
&& - \sum_{a<b} \frac{\rho_{ab}}{2} 
\left[\rule{0mm}{1.5em} n_a \nabla n_b - n_b \nabla n_a\right]^2 
\; + \; \frac{1}{2} g (n_2^2 + n_3^2 + n_4^2), 
\end{eqnarray}
where the indices run from 1 through 5.

The generalized susceptibilities, $\chi$, fall in three groups:
$\chi_c \equiv \chi_{15}$ connecting the charge sector $\{n_1,n_5\}$
with itself, $\chi_s \equiv \chi_{23} = \chi_{24} = \chi_{34}$ 
connecting the spin sector $\{n_2,n_3,n_4\}$ with itself, and
$\chi_{\pi} \equiv \chi_{1(2,3,4)} = \chi_{(2,3,4)5}$ connecting the
spin sector with the charge sector. Similarly with the stiffnesses: 
$\rho_c \equiv \rho_{15}$, $\rho_s \equiv \rho_{23} = \rho_{24} =
\rho_{34}$, and $\rho_{\pi} \equiv \rho_{1(2,3,4)} = \rho_{(2,3,4)5}$.
Below, based on experimental data,
we find that $\rho_s \approx \rho_c$ and $\chi_s \approx
\chi_c$. It is a remarkable fact that the dynamics in two such
distinct sectors as the charge and the spin sector are governed by
coupling strengths of the same size, and it can be taken as one strong
indication of the near $SO(5)$ symmetry of the cuprates. At present
the values of $\chi_{\pi}$ and $\rho_{\pi}$ are not known
experimentally, and it is part of our work to establish a method to
measure them. If the corresponding coupling strengths are the same in
all sectors we denote it the isotropic case, otherwise the anisotropic
case. 

In the following we estimate on a 25~\% accuracy level the typical
zero temperature values of the parameters of the $SO(5)$ model
obtained for various cuprates of the form YBa$_2$Cu$_3$O$_{6+x}$
(YBCO) and La$_{2-x}$Sr$_x$CuO$_4$ (LSCO) with a range of doping
levels $x$. All numerical values are listed in 
Table~\ref{tab:parameters}. First we note that for both materials the
Cu-O-Cu distance \cite{dagotto} is $a = 3.8$~\AA. However, YBCO
contains two CuO planes over a distance of 11.4~\AA\ (the sum of the
alternating inter-layer distances 3.2~\AA\ and 8.2~\AA), while LSCO
contains one CuO plane every 6.6~\AA. Hence, when needed for
normalization purposes we employ $c=6.1$~\AA\ as the typical
inter-layer distance.

From the upper critical field $H_{c2}$ the typical correlation length 
is found to be $\xi \simeq 16$~\AA, \cite{corr} while many
different methods like muon spin rotation, magnetic torque,
magnetization, and kinetic inductance (see Ref.~\onlinecite{pumpin}
and references therein) all result in a London length,
$\lambda_L \simeq  1350$~\AA. Thus the Ginzburg-Landau
parameter is $\kappa = \lambda_L/\xi\simeq 84$.

The connection between the generalized coefficients and the
directly measurable parameters are given below. In the completely
isotropic case where all generalized coefficients are equal we have:
\cite{arovas} 
\begin{eqnarray} 
\tilde{g} &=& g - \chi(2\mu/\hbar)^2, \label{eq:g_tilde} \\
\xi &=& \sqrt{\rho/(-\tilde g)},       \label{eq:xi} \\
\lambda_L &=& \frac{\hbar}{2e} \sqrt{c/\rho \mu_0}. \label{eq:London}
\end{eqnarray}

To estimate $\rho_s$ and $\chi_s$ of the spin sector, experimental
measurements are combined with theoretical calculations of spin waves
within the two dimensional spin 1/2 quantum Heisenberg model of
antiferromagnetism. \cite{chakravarty,manousakis,igarashi} 
The bare coupling constant $J$ is related to $\rho_s$, $\chi_s$ and
the spin wave velocity $v_s$ as \cite{igarashi}

\begin{eqnarray}
\rho_s &=& Z_{\rho}J/4                                  \label{eq:rho_s}\\
\chi_s &=& Z_{\chi}p_a^2/8J                             \label{eq:chi_s}\\
v_s &=& Z_c \sqrt{\rho/\chi} = Z_c \sqrt{2} J/p_a,      \label{eq:v_s}\\
Z_{\rho} &=& 0.72, \quad Z_{\chi}=0.51, \quad Z_c=1.18, \label{eq:ZZZ}
\end{eqnarray}
where for brevity a momentum $p_a \equiv \hbar/a$ has been introduced,
and where the $Z$'s are renormalization constants, which for classical
spin waves all equals 1, but differs from 1 when quantum fluctuations
and spin wave interactions are taken into account. Neutron scattering
experiments on LSCO \cite{hayden91} have led to a determination of
$v_s$ and from that to $J=132$~meV in agreement with other
experiments. From $J$ one calculates $\rho_s = 24$~meV. Independently,
$\rho_s$ have been determined by neutron scattering measurement
\cite{keimer} of the antiferromagnetic correlation length
$\xi_{AFM}(T) \propto \exp(2\pi \rho_s/k_B T)$ also leading to 
$\rho_s = 24$~meV. From Eq.~(\ref{eq:chi_s}) one finds $\chi_s/p_a^2 =
0.49$~eV$^{-1}$.  A more recent neutron scattering experiment on LSCO 
\cite{hayden96} yielded the consistent result $Z_{\chi}=0.4\pm0.1$ and
$J=125$~meV.

In the charge sector we have $\rho_c = (c/\mu_0)(\hbar/2e
\lambda_L)^2$, while $\chi_c = \rho_c/v_c^2$, where $v_c$ is the
sound velocity of the electron liquid. The charge stiffness is readily
obtained from the London length, $\rho_c = (\hbar/2e)^2 (c/\mu_0)
(1/\lambda_L)^2 \approx 18$~meV. The susceptibility $\chi_c$ is
obtained as follows. In ordinary superconductors the
Goldstone mode corresponding to the sound velocity is rendered massive
by the Anderson-Higgs mechanism and turned into a plasmon mode
\cite{pwa63}, but as a theoretical concept it can be calculated, and
it is found to be of the order of the Fermi velocity, $v_F$. Detailed
studies \cite{dagotto,gan} have shown that the dispersion relation for
quasi particles moving around on the lattice of the $t$-$J$ model at
low doping near half filling is governed not by $t$ but is renormalized
$t^* \approx J$. Using the dispersion relation (3.15) of
Ref.~\onlinecite{dagotto},
$\varepsilon_{\bf k} = {\rm const} + J \cos (k_xa) \cos(k_ya) +
\frac{1}{3} J [\cos(2k_xa) + \cos(2k_ya)]$,
 yields $v_c \approx v_F \approx 1.58
J/p_a$. From this we find $\chi_c/p_a^2  \approx  0.42$~eV$^{-1}$.

\begin{table}
\begin{tabular}{r@{$\:$}lr@{$\:$}lr@{$\:$}l}
$a$ & $=3.8$~\AA &  
$\pi \xi^2/a^2$ & $\simeq 56$ &
$c$ & $=6.1$~\AA \\
$\xi$ & $\simeq 16$~\AA & $\lambda_L$ & $\simeq 1350$~\AA &
$\kappa$ & $\simeq 84$ \\
$J$ & $\simeq 0.130$~eV & 
$g$            & $\simeq 9$~meV/\AA$^2$ &
$\tilde{g}$    & $\simeq -53 \; \mu$eV/\AA$^2$ \\
$\rho_s $ & $\simeq  0.024$~eV &
$\chi_s/p_a^2$ & $\simeq 0.49$~eV$^{-1}$ &
$\delta_{\rho}$ & $>-1$  \\ 
$\rho_c$ & $\simeq  0.018$~eV &
$\chi_c/p_a^2$ & $\simeq 0.42$~eV$^{-1}$  &
$\delta_{\chi}$ & $>-0.014$\\
\end{tabular}
\caption{ \label{tab:parameters}
The average values of typical parameters of the $SO(5)$ model based on
experimental data for YBCO and LSCO as described in the text. 
The Cu-O-Cu distance is denoted $a$, while the average distance
between the CuO planes is denoted $c$. For brevity a momentum $p_a
\equiv \hbar/a$ has been introduced. The number of spins in a given
vortex core is estimated by $\pi \xi^2/a^2$. The anisotropy
parameters $\delta_{\rho}$ and $\delta_{\chi}$ are defined in
Sec.~\protect\ref{sec:anisotropic}.
}
\end{table}

To estimate the value of the phenomenological symmetry breaking
constant $g$ in Eq.~(\ref{eq:L_of_n}) we consider complete
antiferromagnetic ordering, {\it i.e.} 
$|{\bf m}|^2 = 1$. In this case the
ordering energy density in the $SO(5)$ model is simply $-\frac{1}{2}g
|{\bf m}|^2 = -\frac{1}{2}g$. On the other hand this energy density
can also be expressed within the $t$-$J$ model as $-J/(2a^2)$, and
therefore $g = J/a^2 \simeq 9$~meV$\:$\AA$^{-2}$. The effective
coupling constant $\tilde{g}$ of Eq.~(\ref{eq:g_tilde}) is much
smaller. Anticipating the discussion in Sec.~\ref{sec:isotropic} of
the $\pi$ resonance frequency $\omega_{\pi} \approx 41$~meV we find the
following estimate: $\tilde{g} = -\chi_{\pi} \omega_{\pi}^2 \simeq -53 \;
\mu$eV/\AA$^2$.

\section{Vortices and core excitations in the $SO(5)$ model}
\label{sec:coreexci}

In Ref.~\onlinecite{arovas} the vortex solutions to the isotropic
$SO(5)$ model have been studied in great detail. To study the
anisotropic case we are going to use a different method. To establish
our method and notation, we will start out in Sec~\ref{sec:isotropic}
by deriving some of the known results for the isotropic case, before
in Sec~\ref{sec:anisotropic} we continue with the anisotropic case.

\subsection{Vortices and resonances in the isotropic case}
\label{sec:isotropic}

In the symmetric version of the $SO(5)$ model the generalized
susceptibilities and stiffnesses are isotropic in superspin space, and
the only symmetry breaking terms are quadratic terms governed by the
chemical potential $\mu$ and the phenomenological constant $g$ chosen
such that superconductivity is favored. The external electromagnetic
fields will now by included through the vector potential $\bf
A$. However, in this section we only keep the interaction with the
$\psi$ part of the order parameter this being the dominating part of
the external fields. In Sec.~\ref{sec:external_field} we will
include the Zeeman interaction between $\bf A$ and the $\bf m$ part of
the order parameter and demonstrate explicitly that this only leads to
minor changes. In this approximation the Lagrangian 
${\cal L}_{\rm iso}$ then has the form:  
 
\begin{eqnarray}
{\cal L}_{\rm iso} =&& 
\frac{1}{2} \chi \left |\partial_t\psi\right|^2 + 
\frac{1}{2} \chi \left |\partial_t{\bf m}\right|^2 + \nonumber\\
&-& \frac{1}{2} \rho 
\left|\left(\nabla + \frac{i2e}{\hbar} {\bf A}\right)\psi\right|^2
- \frac{1}{2}\rho|\nabla{\bf m} |^2 \nonumber\\
&-& \frac{1}{2}\chi (\frac{2\mu}{\hbar})^2|\psi|^2 
+ \frac{1}{2} g{\bf m}^2 
+ \frac{1}{2}\lambda (1-|\psi|^2 - {\bf m}^2) \nonumber\\
&+&\frac{c}{2\mu_0}\left(\frac{1}{c_0^2}|\partial_t{\bf A}|^2 - 
|\nabla\times{\bf A}|^2\right),
\label{eq:L_iso}
\end{eqnarray}
where $c$ is the lattice constant perpendicular to the copper-oxide
planes and $c_0$ the speed of light. 
For later convenience we have incorporated the constraint
Eq.~(\ref{eq:n_vector}) through the Lagrange multiplier $\lambda$. 
Using dimensionless polar coordinates $(s,\phi)$, with $s \equiv
r/\xi$, centred at the origin of the vortex core, we seek solutions of
the form

\begin{equation} \label{eq:psi_m_A}
\psi(s,\phi) = f(s) e^{-i\phi}, \quad
{\bf m} = m(s) {\bf e}_{\beta}, \quad
{\bf A} = \frac{\hbar}{2e \xi} \frac{\alpha(s)}{s} {\bf e}_{\phi},
\end{equation} 
where ${\bf e}_{\beta}$ is an arbitrary unit vector in
$(n_2,n_3,n_4)$-space (equivalent to real space) taken to be $(0,1,0)$
in the following, while ${\bf e}_{\phi}$ is the azimuthal unit
vector. The Euler-Lagrange equation for $\lambda$ yields the constraint

\begin{equation} \label{eq:psi_m_constraint}
f(s)^2 + {\bf m}(s)^2 = 1,
\end{equation}
such that once $f(s)$ is determined so is $m(s)$.
The Euler-Lagrange equation for $\bf m$ is used to express
$\lambda(s)$ in terms of $m(s)$:

\begin{equation} \label{eq:EL_m}
\frac{1}{m}\nabla^2 m = \frac{(\lambda -g)\xi^2}{\rho}.
\end{equation} 
Eqs.~(\ref{eq:psi_m_constraint}) and~(\ref{eq:EL_m}) are then used to
eliminate $m(s)$ and $\lambda(s)$ in the Euler-Lagrange equation for
$f(s)$, and as in Ref.~\onlinecite{arovas} we end up with:

\begin{equation} \label{eq:EL_f}
\nabla^2 f + 
\frac{f}{1-f^2} (\partial_s f)^2 + f (1-f^2)
\left[1-(\frac{\alpha-1}{s})^2\right] = 0.
\end{equation}
The Euler-Lagrange equation for $\alpha(s)$ becomes

\begin{equation} \label{eq:EL_alpha}
\partial_s^2 \alpha - \frac{1}{s} \partial_s \alpha =
\frac{(\alpha -1)}{\kappa^2}f^2.
\end{equation}
Equations (\ref{eq:EL_f}) and (\ref{eq:EL_alpha}) are solved by the
numerical shooting method \cite{arovas} and yields $f(s)$ and
$\alpha(s)$. 

Introducing the ``effective potential'' $V_0(s)$ as the right hand
side of equation Eq.~(\ref{eq:EL_m}),

\begin{equation} \label{eq:V_iso(s)}
V_0(s) \equiv  \frac{(\lambda -g)\xi^2}{\rho} =
\frac{1}{m}\nabla^2 m = 
- \frac{f \nabla^2 f}{(1-f^2)} - \frac{(\partial_s f)^2}{(1 - f^2)^2},
\end{equation}
we can use the solution of $f(s)$ to determine $V_0(s)$, which
in turn results in the Euler-Lagrange equation for $\bf m$ of the form

\begin{equation} \label{eq:zero_mode}
\left[-\nabla^2 + V_0(s)\right] \: m = 0.
\end{equation}
Naturally, by construction, Eq.~(\ref{eq:zero_mode}), is automatically
fulfilled using $m(s) = \sqrt{1-f(s)^2}$, but for the forthcoming
studies of core excitations it is useful to think of the static core
as corresponding to the zero energy solution of the
Schr\"{o}dinger-like equation Eq.~(\ref{eq:zero_mode}) where 
$V_0(s)$ clearly plays the role of an effective potential.
For the same reasons we rewrite Eq.~(\ref{eq:EL_f}) for $f(s)$ by the
use of Eq.~(\ref{eq:V_iso(s)}) for $V_0(s)$:
\begin{equation} \label{eq:pi_mode}
\left[-\nabla^2 + \frac{1}{s^2} + V_0(s)
+\frac{\alpha(2-\alpha)}{s^2} \right] f = f.
\end{equation}

Thus having established the notation and found the static vortex
solutions we now turn to the problem of finding excitations in the
vortex core. We denote the static vortex solution by ${\bf n}_0$,
and due to the $SO(3)$ symmetry in the spin sector we are free
to choose the direction of ${\bf n}_0$ at will. In anticipation of the
treatment in Sec.~\ref{sec:external_field} where an external magnetic
field in the $z$ direction forces ${\bf n}_0$ to lie in the $xy$ plane
we choose ${\bf n}_0 \propto {\bf e}_y$, {\it i.e.} only the second of
the three ${\bf n}_0$ components is nonzero: 

\begin{equation} \label{eq:n0}
{\bf n}_0 = \{ f(s) \cos(\phi), 0, m(s), 0, -f(s) \sin(\phi) \}.
\end{equation}
We seek excitations $\delta {\bf n}$ which are of lowest order in the
deviations $\delta \psi$ and $\delta {\bf m}$. These turns out to be
perpendicular to ${\bf n}_0$, {\it i.e.} $\delta {\bf n} \propto {\bf
e}_x$ or $\delta {\bf n} \propto {\bf e}_z$ or linear combinations
thereof, and hence of the form

\begin{equation} \label{eq:dn}
\begin{array}{rcl}
\delta {\bf n}_x & = & \{ 0, 1, 0, 0, 0 \} \: \delta m, \\
\delta {\bf n}_z & = & \{ 0, 0, 0, 1, 0 \} \: \delta m, \\
\delta m         & = & \delta m_{nl}(s) e^{i l \phi} e^{-i\omega t}.
\end{array}
\end{equation}
Throughout this work we are dealing with cylindrical symmetric
vortices, so the excitations are characterized by the angular momentum
$l$ and the radial index $n$. Of course, $\delta {\bf n}$ is not a
complex vector, so the notation 
$\exp(il \phi)$ is merely a short hand notation for either 
$\cos(l \phi)$ or $\sin(l \phi)$. The Lagrangian is now written to
second order in $\delta {\bf n}$ as 
${\cal L} = {\cal L}^{(0)}_{\rm iso}({\bf n}_0) + 
{\cal L}^{(2)}(\delta {\bf n})$. The explicit form of the second order
term is

\begin{equation} \label{eq:L2}
{\cal L}^{(2)}(\delta {\bf n}) = 
\frac{1}{2} \chi |\partial_t \delta {\bf n}|^2 -
\frac{1}{2} \rho |\nabla \delta {\bf n}|^2 +
\frac{1}{2} [g - \lambda(s)] \: |\delta {\bf n}|^2.
\end{equation}
Assuming solutions $\delta {\bf n}$ of the form given in
Eq.~(\ref{eq:dn}) the Euler-Lagrange equation for $\delta m_{nl}(s)$
then takes the form of the following eigenvalue equation:

\begin{eqnarray} \label{eq:dm}
\left[ -\nabla^2 + \frac{l^2}{s^2} + V_0(s) 
\right] \delta m_{nl} = \varepsilon \delta m_{nl}, \\
\varepsilon = \frac{\chi \xi^2}{\rho} \omega^2 =
              \frac{\chi}{-\tilde{g}} \omega^2. \nonumber
\end{eqnarray}
Using the approximate $SO(5)$ symmetry we can immediately find two
analytical solutions $\delta m_{00}$ and $\delta m_{01}$ to
Eq.~(\ref{eq:dm}). Due to the exact spin rotation symmetry it does not
cost any energy to rotate the order parameter ${\bf n}_0$ of
Eq.~(\ref{eq:n0}) in spin space. Rotating ${\bf n}_0$ a small angle
$\delta \theta$ 
in the $(n_3,n_4)$ plane produces the deviation $\delta {\bf n}_{00} =
\{0,0,0,\delta \theta \: m(s),0\}$. If the trial solution $\delta
m_{00}(s) \propto m(s)$ is used (note that $l =0$), we see from
Eq.~(\ref{eq:zero_mode}) that as expected Eq.~(\ref{eq:dm}) is
satisfied with $\varepsilon = 0$. Thus in the the effective potential
description the ground state vortex configuration corresponds to a
zero energy and zero angular momentum mode. 

The second solution is found by rotating between the charge sector and
the spin sector. In a perfect $SO(5)$ symmetric model such a rotation
does not cost any energy. However, one central idea in the $SO(5)$
model of high $T_c$ superconductors is that the $SO(5)$ symmetry is
only approximate. It costs a finite energy to rotate between the spin
and charge sectors. Experimentally this is reflected by the 41~meV
$\pi$ resonance, and theoretically by the symmetry breaking term
$\frac{1}{2} g {\bf m}^2$ of Eq.~(\ref{eq:L_of_n}). We thus expect
that by rotating ${\bf n}_0$ a small angle $\delta \theta$
in the $(n_1,n_4)$ plane a deviation $\delta {\bf n}_{01} =
\{ 0,0,0,\delta \theta \: f(s) \cos(\phi),0 \} $ is produced which is
an eigen-excitation with $\varepsilon > 0$. If the trial solution
$\delta m_{01}(s) \propto f(s) \cos(\phi)$ is inserted into
Eq.~(\ref{eq:dm}) (note that $l =1$), we obtain

\begin{equation} \label{eq:dm01}
\left[-\nabla^2 + \frac{1}{s^2} + V_0(s)\right] f 
= \varepsilon f.
\end{equation}
It is seen from Eq.~(\ref{eq:pi_mode}) that this equation is indeed
satisfied by $f(s)$ if $\varepsilon=1$ and if the additional potential
$\alpha(2-\alpha)/s^2$ can be neglected. It turns out that the large
value of the Ginzburg-Landau parameter, $\kappa \simeq 84$, indeed does
make the additional potential negligible. Numerical calculations show
$\alpha(2-\alpha)/s^2 < 0.001$ (for any value of $s$) which is
much smaller than  $V_0(s) \simeq 1$ and $\varepsilon =
1$. The approximate 
nature of the solution is not surprising, since the external magnetic
field does in fact break the $SO(5)$ symmetry by coupling only to the
$(n_1,n_5)$ components of the order parameter. However, the larger a
value of $\kappa$ the weaker this symmetry breaking appears, and in
the limit of infinite $\kappa$ the approximate solution becomes
exact. Since the ground state energy is set to be zero as the zero
energy mode, the excitation energy of the $\delta m_{01}$ resonance is
given by $\varepsilon = 1$, or going back to frequency: $\omega_{\pi} =
\sqrt{-\tilde{g}/\chi}$, which is in accordance with that of the $\pi$
resonance given in Ref.~\onlinecite{zhang}.

In conclusion we note that both $\delta m_{00}$ and $\delta m_{01}$
contains no radial nodes, hence the notation $n=0$. Any other
excitation or resonance would contain more nodes and thus have higher
energies. Since $\delta m_{01}$ corresponds to a resonance at the
bottom edge of the continuum we can infer that for the isotropic case no
bound collective excitations exist in the antiferromagnetic vortex core.

\subsection{Vortices and excitations in the anisotropic case}
\label{sec:anisotropic}

We now turn to the anisotropic case. As discussed in
Sec.~\ref{sec:SO5} $\rho_s \approx \rho_c$ and $\chi_s 
\approx \chi_c$. In the following we therefore study the consequences
of anisotropies arising from $\rho_{\pi} \neq \rho_s = \rho_c$ and
$\chi_{\pi} \neq \chi_s = \chi_c$:

\begin{equation} \label{eq:asym_notation}
\begin{array} {rclrcl}
\rho          & \equiv & \rho_s = \rho_c, & \qquad \qquad
\chi          & \equiv & \chi_s = \chi_c, \\
\rho_{\pi}    & \equiv & \rho + \Delta \rho,& \qquad \qquad
\chi_{\pi}    & \equiv & \chi + \Delta \chi,\\
\delta_{\rho} & \equiv & \Delta \rho/\rho, & \qquad \qquad
\delta_{\chi} & \equiv & \Delta \chi/\chi.
\end{array}
\end{equation}
These anisotropies are not known
experimentally, and it is part of our work to establish a method to
measure it. We begin by finding constraints on them. The stiffness has
to be a positive number, {\it i.e.} $\rho_{\pi} = \rho
(1+\delta_{\rho}) > 0$ or $\delta_{\rho} > -1$. The $\pi$
susceptibility $\chi_{\pi}$ is related to the $\pi$ resonance and to
the coupling strengths $g$ and $\tilde{g}$. Defining $\gamma \equiv
(g-\tilde{g}_0)/g$, where the subscript 0 refer to the isotropic
case, we can estimate $\gamma$ using the values listed in
Table~\ref{tab:parameters} and find $\gamma \simeq 1.014$. The
relation between $\tilde{g}$ and $\tilde{g}_0$ can be written as

\begin{equation} \label{eq:gpi_tilde}
\tilde{g} = 
g-\chi_{\pi}(\frac{2\mu}{\hbar})^2 = 
\tilde{g}_0 \: \frac{1-(1+\delta_{\chi})\gamma}{1-\gamma}.
\end{equation}
To ensure the superconducting phase it is mandatory to have $\tilde{g}
< 0$ and hence from the enumerator in Eq.~(\ref{eq:gpi_tilde}) that
$1-(1+\delta_{\chi})\gamma < 0$ or $-0.014 < \delta_{\chi}$. This
constraint is listed in Table~\ref{tab:parameters}.

\begin{figure}[h]
\epsfxsize=\columnwidth\epsfbox{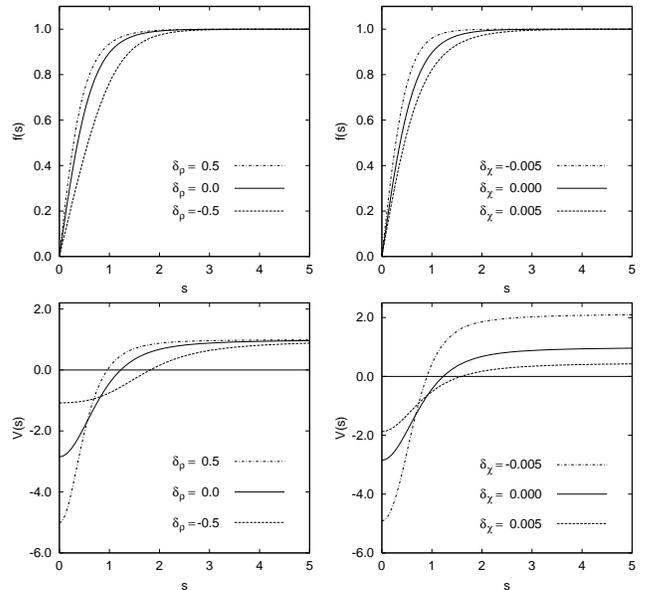}
\caption{\label{fig:vortex_cores}
A display of the vortex cores for various values of the anisotropy
parameters. Graphs of the modulus $f(s)$ of the superconducting
order parameter as well as of the effective potential $V(s)$. The
parameters are $\kappa=84$ and $\gamma = 1.014$. When
$\delta_{\rho}$ is varied, $\delta_{\chi}=0$ and vice versa.
}
\end{figure}

The Lagrangian ${\cal L}_{\rm ani} = {\cal L}_{\rm iso} + \Delta
{\cal L}$ in the anisotropic case differs from the Lagrangian ${\cal
L}_{\rm iso}$ in the isotropic case by $\Delta {\cal L}$ containing
terms proportional to the anisotropies $\delta_{\rho}$ and
$\delta_{\chi}$.  
The Euler-Lagrange equations for $f$, $m$, and $\alpha$ corresponding
to ${\cal L}_{\rm ani}$ become

\begin{eqnarray} 
\nabla^2 f &+& 
\frac{f (\partial_s f)^2}{1 -f^2} + 
\left[1 - (\frac{\alpha - 1}{s})^2 \right] f(1-f^2) \\ 
\label{eq:f_ani}
&=& \left[ 
\frac{2\delta_{\chi}}{1+\delta_{\chi}}(1-\frac{g}{\tilde{g}})- 
\frac{2\delta_{\rho}}{1+\delta_{\rho}}\left(\frac{\alpha-1}{s}\right)^2 
\right] \! f^3 (1\!-\!f^2), \nonumber \\ 
\label{eq:m_ani}
\frac{\nabla^2 m}{m}  &=& 
\frac{(\lambda -g) \xi^2}{\rho_{\pi}}
+ \frac{\delta_{\rho}}{1 + \delta_{\rho}}
\left[(\nabla m)^2 + m \nabla^2 m \right],\\
\label{eq:a_ani}
\partial_s^2 \alpha  &=& 
\frac{1}{s} \partial_s \alpha + \frac{\alpha - 1}{\kappa^2}f^2 
\left[1 - \frac{\delta_{\rho}}{1 + \delta_{\rho}}f^2 \right].
\end{eqnarray}
It is seen how the anisotropy leads to more nonlinear terms in the
differential equations. Using the same numerical methods as in
Sec.~\ref{sec:isotropic} we study the static vortex cores for various
values of $\delta_{\rho}$ and $\delta_{\chi}$. Some results are shown
in Fig.~\ref{fig:vortex_cores}. 

In the isotropic case the static vortex core led to an effective
potential description with $V_0(s)$ defined as the right hand side of 
Eq.~(\ref{eq:EL_m}). Similarly, for the anisotropic case we now define
an effective potential $V(s)$ as the right hand side of
Eq.~(\ref{eq:m_ani}) which then satisfies the equation

\begin{equation} \label{eq:V_ani}
[-\nabla^2  + V(s)] \: m = 0.
\end{equation}
$V(s)$ can be expressed in terms of $f$ by using $m^2 = 1 - f^2$, 
and thus it can be found by solving Eqs.~(\ref{eq:f_ani})
and~(\ref{eq:a_ani}). Due to the anisotropy the effective potential
and hence the excitation spectrum changes. The effective potential for
some anisotropy parameters are shown in Fig.~\ref{fig:vortex_cores}.
We study the transverse excitations given by Eq.~(\ref{eq:dn}) and
find the following eigenvalue equation to be fulfilled by $\delta
m_{nl}(s)$:

\begin{equation} \label{eq:dm_ani}
[-\nabla^2 + \frac{l^2}{s^2} + V(s)] \: \delta m_{nl} = 
\frac{\xi^2_{\pi} \chi_{\pi}}{\rho_{\pi}} \omega^2 \delta m_{nl} =
\varepsilon \: \delta m_{nl}.
\end{equation}

\begin{figure}[h]
\epsfxsize=\columnwidth\epsfbox{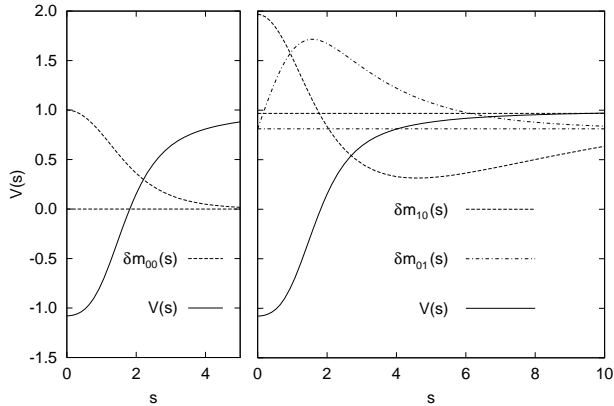}
\caption{\label{fig:bound_exci}
The radial part $\delta m_{nl}$ of bound antiferromagnetic excitations
in the vortex core in an anisotropic case. To show the effects more
clearly we have chosen the rather extreme parameter value
$\delta_{\rho} = -0.5$, while having $\kappa=84$, $\delta_{\chi} = 0$,
and $\gamma = 1.014$. To the left is shown the zero energy state 
$\delta m_{00}$ corresponding to the static vortex core. To the right
are shown the first two excitations $\delta m_{01}$ and $\delta
m_{10}$ above the zero energy mode. The full line in both panels is
the effective potential $V(s)$. The ordinate axis accounts
for $V(s)$ in the energy units of
Eq.~(\protect\ref{eq:dm_ani}). The excitations $\delta m_{nl}$ are
given in arbitrary units. The eigenenergies are represented by the
dashed horizontal lines with $\varepsilon_{00} = 0.00$,
$\varepsilon_{01} = 0.81$, and $\varepsilon_{10} = 0.97$. 
} 
\end{figure}

As in the isotropic case a zero energy solution corresponding to the
static vortex solution is trivially given. What is new in the
anisotropic case is that now bound excitations do exist with
$\varepsilon < V(\infty)$. Some of them are shown in
Fig.~\ref{fig:bound_exci}.

\begin{figure}[h]
\centerline{\epsfxsize=60mm\epsfbox{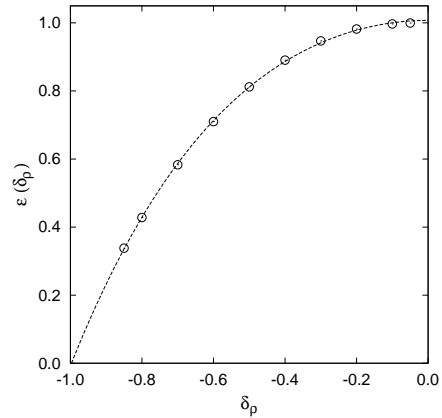}}
\caption{\label{fig:bound_erg}
The energy $\varepsilon$ of the lowest bound excitation as a function
of the anisotropy $\delta_{\rho}$ for $\kappa = 84$, $\delta_{\chi} =
0$, and $\gamma = 1.014$. The dotted line is a fit, 
$\varepsilon(\delta_{\rho}) \simeq 1.0 - 0.7 \delta_{\rho}^2 - 0.3
\delta_{\rho}^4$. Note that $\varepsilon(0) = 1$ and $\varepsilon(-1)
= 0$.
}
\end{figure}

In Fig.~\ref{fig:bound_erg} we plot as a function of the anisotropy
parameter $\delta_{\rho}$ the energy of the lowest
excited state $\delta m_{01}$ above the ever present zero energy mode
$\delta m_{00}$. It is seen how $\delta m_{01}$ evolves from the
scattering resonance, the $\pi$ resonance,  discused in
Sec.~\ref{sec:isotropic} with $\varepsilon = V(\infty)$ at
the isotropic point $\delta_{\rho} = 0$ to a strongly bound state with
$\varepsilon = 0$ at $\delta_{\rho} = -1$, the largest negative value
allowed. At this point the excited state thus coincide energetically
with the zero energy state indicating the phase transition from
superconductivity to antiferromagnetism -- the gap of the
$\pi$ excitations has collapsed.

Our calculations thus lead to the following prediction. For a given
anisotropy $\delta_{\rho} < 0$ neutron scattering will in zero
magnetic field show the $\pi$ resonance with $\varepsilon = V_{\rm
ani}(\infty)$ or frequency $\omega = \omega_{\pi}$. As the magnetic
field is turned on more and more vortices are created. Each of them
supports a $\delta m_{01}$ excitation with an energy $\varepsilon_{01}
< V(\infty)$ or frequency $\omega_{01} <
\omega_{\pi}$. Consequently an excitation resonance should show up at the
low-energy side of the $\pi$ resonance, and the amplitude of this
resonance scales with the magnetic field. 

Similar results does not hold for $\delta_{\chi} \neq 0$. The reason
for this can be traced back to the behavior of $V$ shown in
Fig.~\ref{fig:vortex_cores}. When $\delta_{\chi}$ is made positive the
potential widens as in the case of negative $\delta_{\rho}$, however,
at the same time the asymptotic value drops. As a result no
excitations are bound in the potential. In the case of negative
$\delta_{\chi}$ the potential narrows down, but although the
asymptotic value rises, it does not rise enough to bind any
excitations. 

It should be added that this change in the asymptotic
value of $V$ due to $\delta \chi$ is in accordance with the 
result for the $\pi$ resonance in a bulk superconductor without the
presense of vortices. In the dimensionless units of the problem, see
Eq.~(\ref{eq:dm}), the frequency $\omega_{\pi}$ of the $\pi$ resonance
is given by 

\begin{equation} \label{eq:omega_pi}
\frac{\chi_{\pi}}{-\tilde{g}} \omega_{\pi}^2 =
1 + 2\frac{\Delta \chi}{\tilde{g}} \frac{(2\mu)^2}{\hbar^2} = 
\frac{(1-\gamma)+\delta_{\chi}\gamma}{(1-\gamma)-\delta_{\chi}\gamma}.
\end{equation}
Inserting the values $\delta_{\chi} = -0.005$, 0, and 0.005 we find 
2.14, 1.00, and 0.47, respectively, as $V(\infty)$ in
Fig.~\ref{fig:vortex_cores}. 

\section{Goldstone modes}
\label{sec:Goldstone}

In Sec.~\ref{sec:coreexci} we studied the zero energy modes and the
excitations of the antiferromagnetic order parameter $\bf m$ in the
superconducting vortex core. The zero energy modes $\delta {\bf n}_x$
and $\delta {\bf n}_z$ in Eq.~(\ref{eq:dn}) were found to be
degenerate gapless Goldstone modes. This degeneracy is a result of two
approximations, one being the neglect of the Zeeman interaction between 
$\bf m$ and the external magnetic field $\bf B$, the other being the
neglect of inter-layer interaction between the spins. In this section
we show how the degeneracies are lifted and massive Goldstone modes
appear when the approximations are abandoned.

\subsection{Coupling of core excitations to the external magnetic field}
\label{sec:external_field}

The primary effect of the external magnetic field 
${\bf B} = B{\bf e}_z$ on the
system is the creation of superconducting vortices through the
interaction with the superconducting order parameter $\psi$. To a good
approximation the antiferromagnetic core described by $\bf m$ can be
treated disregarding the Zeeman coupling between $\bf B$ and $\bf
m$. We now take this coupling into account. Using standard field
theoretic methods \cite{fradkin} the spin operator $\hat{\bf S}_j$
acting on site $j$ can be expressed by a classical field ${\bf h}_j$,

\begin{equation} \label{eq:Sj}
\hat{\bf S}_j \; \longrightarrow \; 
s \: e^{i{\bf Q} \cdot {\bf R}_j} {\bf h}_j,
\end{equation}
where  $s=1/2$ and ${\bf Q} = \pi(a^{-1}, a^{-1}, c^{-1})$ is
the antiferromagnetic ordering vector. We then have 
$\exp[i{\bf Q} \cdot {\bf R}_j] = (-1)^j$, {\it i.e.} 1 on sublattice
A and $-1$ on sublattice B. Since the system is close to be completely
antiferromagnetically ordered ${\bf h}_j$ is written as

\begin{equation} \label{eq:hj}
{\bf h}_j =  {\bf m}_j + (-1)^j  {\bf l}_j,
\end{equation}
where ${\bf l}_j$ denotes a small ferromagnetic component on top of
the antiferromagnetic background ${\bf m}_j$. Both ${\bf l}_j$ and
${\bf m}_j$ are slowly varying fields in space. The rapid
antiferromagnetic variation from site to site is explicitly taken into
account by prefactors $(-1)^j$. The smallest deviation possible is
obtained by having ${\bf l}_j$ perpendicular to ${\bf m}_j$, and since
${\bf h}_j$ is normalized to unity, we obtain to lowest order in 
${\bf l}_j$ that 

\begin{equation} \label{eq:l_constraint}
|{\bf h}_j|^2 = 1, \qquad 
{\bf l}_j \!\cdot\! {\bf m}_j = 0, \qquad 
|{\bf m}_j|^2 = 1.
\end{equation}
The Hamiltonian $\hat{H}_B$ corresponding to the coupling between the
spins and $\bf B$ is

\begin{equation} \label{eq:H_B}
\hat{H}_B = \sum_j g^*\mu_B \hat{\bf S}_j \!\cdot\! {\bf B}
\;\longrightarrow\;
\frac{1}{2} g^*\mu_B {\bf B} \!\cdot\! \sum_j (-1)^j {\bf h}_j.
\end{equation}
Only the ferromagnetic component yields a non-zero contribution to the
sum, and after taking the continuum limit we end with a Lagrangian
density ${\cal L}_B$ given by

\begin{equation} \label{eq:L_B}
{\cal L}_B = -\frac{1}{2} \frac{g^*\mu_B}{a^2} 
             {\bf B} \!\cdot\! {\bf l}({\bf r}).
\end{equation}

The appearance of a non-zero ferromagnetic component 
${\bf l}({\bf r})$ leads to a loss of antiferromagnetic ordering
energy. This has to be included in the description through the
Hamiltonian $\hat{H}_{\rm AFM}$ describing the spin-spin interaction:

\begin{equation} \label{eq:H_AFM}
\hat{H}_{\rm AFM} = 
J \sum_{\langle j,j' \rangle} \hat{\bf S}_j \!\cdot\! \hat{\bf S}_{j'}
\; \longrightarrow \;
 -\frac{1}{4}J\sum_{\langle j,j'\rangle}{\bf h}_j\!\cdot\!{\bf h}_{j'}.
\end{equation}
Performing the sum and taking the continuum limit results in the
Lagrangian density ${\cal L}_{\rm AFM}$,

\begin{equation} \label{eq:L_AFM}
{\cal L}_{\rm AFM} = \frac{1}{2} \frac{J}{a^2}
                     \left[1 - {\bf l}({\bf r})^2 \right]. 
\end{equation}

The constraint ${\bf l}_j \cdot {\bf m}_j = 0$ is incorporated
through a Lagrange multiplier $\lambda'$ and we end with the
following Lagrangian density ${\cal L}'$ of the magnetic
effects 

\begin{equation} \label{eq:Lp}
{\cal L}' =
-\frac{1}{2} \frac{J}{a^2} {\bf l}({\bf r})^2
+\left(\lambda'{\bf m}-\frac{1}{2} \frac{g^*\mu_B}{a^2}{\bf B}\right)
 \!\cdot\! {\bf l}({\bf r}) + \frac{1}{2} \frac{J}{a^2}.
\end{equation}

An effective Lagrangian ${\cal L}_{\rm mB}({\bf m})$ for the
interaction between $\bf m$ and $\bf B$ is found from the partition
function $Z = 
\int {\cal D} {\bf m} \int {\cal D} {\bf l} \int {\cal D} \lambda'
\exp[iS({\bf m},{\bf l},\lambda')/\hbar] =
\int {\cal D} {\bf m} \exp[iS_{\rm eff}({\bf m})/\hbar]$ from which the
final result can be extracted after integrating out ${\bf l}$ and
$\lambda'$: 

\begin{equation} \label{eq:L_mag}
{\cal L}_{\rm mB}({\bf m}) =
-\frac{1}{2} \Gamma ({\bf B} \!\cdot\! {\bf m})^2
+\frac{1}{2} \Gamma {\bf B}^2 + \frac{1}{2} \frac{J}{a^2},
\end{equation}
where $\Gamma = (g^* \mu_B)^2/(4Ja^2)$. Adding $\cal L_{\rm mB}$ to
the Lagrangian of the anisotropic case $\cal L_{\rm ani}$ leads to the
following eigenvalue equation for an excitation $\delta {\bf
n}_{\beta} = \delta m {\bf e}_{\beta}$, with $\beta = x$ or $z$,

\begin{equation} \label{eq:dm_mag}
[-\nabla^2 + V(s)] \delta m = 
\frac{\xi^2_{\pi}\chi_{\pi}}{\rho_{\pi}}   
\left[\omega^2 - 
      \frac{\Gamma}{\chi_{\pi}}B^2{\bf e}_z \!\cdot\!{\bf e}_{\beta}
\right] \delta m.
\end{equation}
As was the case for Eq.~(\ref{eq:dm}) two eigensolutions proportional
to $m$ can now be found . One mode has $\beta = x$ ({\it i.e.}  
${\bf e}_z \!\cdot\!{\bf e}_{\beta} = 0$) and remain a zero energy
mode: 

\begin{equation} \label{eq:ox}
\hbar \omega_x=0.
\end{equation}
The other mode has $\beta=z$ ({\it i.e.} ${\bf e}_z \!\cdot\!{\bf
e}_{\beta} = 1$), resulting in a nonzero energy, which can be
estimated by setting $B = \Phi_0 / \pi\lambda_L^2$ and using the
parameters of Table~\ref{tab:parameters}: 

\begin{equation} \label{eq:oz}
\hbar \omega_z = \hbar \sqrt{\frac{\Gamma}{\chi_{\pi}}} \: B = 
\frac{\sqrt{2} \hbar^2}{m^* \lambda_L^2} \simeq 5.9 \; \mu{\rm eV}.
\end{equation}
We can thus conclude that the presense of the external magnetic field
in fact does break the degeneracy of the two gapless Goldstone modes,
leaving only one mode gapless while rendering the other massive. 
The rather small value of the gap, 5.9~$\mu$eV $\approx$ 68~mK, would be
very difficult to observe in a neutron scattering experiment. It is
two orders of magnitude smaller than the $\pi$ resonance, and hence it
is seen to be a good approximation to disregard the interaction
between the external magnetic field and the antiferromagnetic order
parameter.

\subsection{Inter-layer coupling}
\label{sec:inter-layer_coupling}

To this point we have only treated one single CuO plane. Naturally, to
stabilize the order parameter, interaction between the layers has
tacitly been assumed. In this section we are explicitly going to
include that part of the inter-layer coupling which arises from the
antiferromagnetic coupling between the CuO planes. We model this
coupling by a Hamiltonian $\hat{H}'$, where spins at site $j$ in the
CuO plane $\zeta$ are interacting with the closest spins in the
neighboring planes $\zeta \pm 1$ as follows: 

\begin{equation} \label{eq:Hp}
\hat{H}' = \frac{1}{2}J' \sum_{j,\zeta} \hat{\bf S}_{j,\zeta} \!\cdot\! 
\left[ \hat{\bf S}_{j,\zeta-1} + \hat{\bf S}_{j,\zeta+1} \right].
\end{equation}
The inter-layer coupling $J'$ is much smaller than the
intralayer coupling $J$, namely $J' \simeq 4\!\times\!10^{-5} J$. 
\cite{cheong} In this
specific model there is no coupling between sites having different
in-plane index $j$, and we are led to consider 1D spin chains
perpendicular to the planes. As before, see Eqs.~(\ref{eq:Sj})
and~(\ref{eq:hj}), the spins are represented by the classical fields
${\bf h}$, ${\bf m}$, and ${\bf l}$:

\begin{equation} \label{eq:Sjz}
\hat{\bf S}_{j,\zeta} \longrightarrow
 s(-1)^{\zeta} {\bf h}_{j,\zeta} , \qquad
{\bf h}_{j,\zeta} = 
{\bf m}_{j,\zeta} + (-1)^{\zeta} {\bf l}_{j,\zeta}.
\end{equation}
Henceforth, we drop the site index $j$ and focus on just one of the
spin chains. The Hamiltonian $H'$ is now expressed in terms of the
classical fields, and in the continuum limit we obtain the form

\begin{equation} \label{eq:Hpcont}
\hat{H}' \longrightarrow  
-\frac{J'}{8 a^2} \int d^2 {\bf r} \sum_{\zeta}
\left\{ 2 - [{\bf h}_{\zeta  }({\bf r}) - 
             {\bf h}_{\zeta-1}({\bf r})]^2 \right\}.
\end{equation}
Defining $\Delta {\bf m}_{\zeta}={\bf m}_{\zeta}-{\bf m}_{\zeta-1}$ 
we obtain the following form of the Lagrange density, ${\cal L}_{\rm
ilc}$ for the inter-layer coupling:

\begin{equation} \label{eq:Lilc}
{\cal L}_{\rm ilc} = -\frac{J'}{8 a^2}  \sum_{\zeta} 
\left[ 2 - \Delta {\bf m}_{\zeta}^2 \right].
\end{equation}

As in Sec.~\ref{sec:external_field}, excitations $\delta {\bf m}$ in
the order parameter ${\bf m} = m {\bf e}_y$ are  sought in the
perpendicular directions ${\bf e}_x$ and ${\bf e}_z$. The motion in
these two directions is independent of each another, and we write the
excitations  $\delta {\bf m}_{\beta}$, $\beta=x$ or $z$,  as

\begin{equation} \label{eq:dm_ilc}
\delta {\bf m}_{\beta}(\zeta) = 
\theta_{\beta}(\zeta) \: {\bf e}_{\beta},
\end{equation}
where the amplitude $\theta_{\beta}(\zeta)$ is $m$ times the (small)
angle by which the order parameter $m$ in plane $\zeta$ is tilted away
from its equilibrium position. The eigenmodes are found by the Fourier
transformation 

\begin{equation} \label{eq:theta_k}
\theta_{\beta}(\zeta) = \sum_k \theta_{k,\beta} e^{i k \zeta c},
\qquad k = n \frac{2 \pi}{N_c c}, \quad n=1,2,3,\ldots,
\end{equation}
where $N_c$ is the number of CuO planes in the sample. Because of the
Fourier transform it is more natural to work with Lagrange functions,
$L = \int {\cal L} d^2 {\bf r}$, rather than Lagrange
densities. However, in the end by dividing $L$ with $N_c \xi^2$, the
number of planes times the effective area of a vortex, the Lagrange
function is rendered into a Lagrange density. The resulting Lagrangian
density ${\cal L}^{(2)}_{\rm ilc}(\delta {\bf m}_{\beta})$ thus becomes

\begin{eqnarray} \label{eq:L2_ilc}
{\cal L}^{(2)}_{\rm ilc}(\delta {\bf m}_{\beta}) &=&
\sum_{k,\beta} \left\{
\frac{\chi}{2}  |\partial_t \theta_{k,\beta}|^2
-\frac{J'}{4a^2} [1\!-\!\cos(kc)] |\theta_{k,\beta}|^2 \right\}
\nonumber \\
&=& \frac{1}{2} \chi \sum_{k,\beta} \left[
|\partial_t \theta_{k,\beta}|^2 -
\omega_{k,\beta}^2 |\theta_{k,\beta}|^2 \right],
\end{eqnarray}
where, after using $\chi = \hbar^2/8Ja^2$, the eigenfrequencies
$\omega_{k,\beta}$ are seen to be

\begin{equation} \label{eq:omega_k}
\hbar \omega_{k,\beta} = 2 \sqrt{J'J [1 - \cos(k c)]}.
\end{equation}
The inter-layer coupling thus splits the $2N_c$-fold degenerate
Goldstone modes in a stack of $N_c$ vortices. The two $k=0$ modes,
$\omega_{0,\beta}=0$, remain zero energy modes. However, since 
$\omega_{k,(x,z)} = \omega_{2\pi-k,(x,z)}$, the rest of the
modes, $k>0$, split up in a quasi continuous band consisting of $N_c/2
-1$ four-fold degenerate massive Goldstone modes. The most massive
modes are found for $k=\pi/c$ with an energy

\begin{equation} \label{eq:emax}
\hbar \omega_{\pi/c,\beta} = \sqrt{8J'J} \simeq 2.2~{\rm meV}
\approx 26~{\rm K}.
\end{equation}
We note that for two-layer compounds such as YBCO the effect of the
inter-layer coupling can be estimated by setting $N_c=2$ in the above
calculation. In this case only $k=0$ and $k=\pi/c$ occurs, and the
degeneracy splitting is given by the stated 2.2~meV or 26~K.
We conclude that also the inter-layer coupling produces
only minor effects in the excitation spectrum as compared to the
$\pi$ excitation, however, the estimated splitting of 2.2~meV
is resolvable with the existing neutron scattering spectrometers. 
For completeness we note that the combined effect
of the external magnetic field and the inter-layer coupling on the
Goldstone modes  is given by $\omega = \sqrt{\omega_\beta^2 +
\omega_{k,\beta}^2 }$.

\section{Conclusions}
\label{sec:conclusions}

The excitations in the antiferromagnetic cores of superconducting
vortices in the $SO(5)$ model have been studied. By examining the
existing literature on experimental results connecting to the values
of the parameters of the model, we have found that the stiffnesses in
the charge and spin sectors are nearly identical, $\rho_c \approx
\rho_s$, and likewise for the susceptibilities, $\chi_c \approx
\chi_s$. This remarkable fact serves as good support of the idea of the
existence of a $SO(5)$ symmetry in the high $T_c$ cuprates. 

We have predicted within the $SO(5)$ model that bound localized
excitations exist when asymmetries arise between the $\pi$ sector of
the parameters and the spin and charge sectors. If they exist, these
excitations could be observed in neutron scattering experiments as
side peaks to the already observed $\pi$ excitation, side peaks with
an amplitude proportional to the number of vortices and thereby
proportional to the applied external magnetic field.  

Finally, we have predicted the splitting of the degenerate zero energy
mode as a function of applied magnetic field and the inter-layer
coupling. The effect of the magnetic field is minute, only a few
$\mu$eV, and thus not possible to detect with present day neutron
scattering technology. The effect of the inter-layer coupling, on the
other hand, is of the order of 2~meV and hence detectable in inelastic
neutron scattering experiments. The test of the existence of these
core excitations would constitute a crucial test of the $SO(5)$ model.
The expected signal should only be present in the superconducting
phase, and it should be proportional with the number of vortices,
{\it i.e.} with the applied magnetic field.

\section{Acknowledgements}
\label{sec:acknowledgements}
We thank Niels Hessel Andersen and Poul Erik Lindelof for useful
discussions concerning the experimental possibilities. H.B.\ is
supported by the Danish Natural Science Research
Council through Ole R\o mer Grant No.\ 9600548.


\begin{references}
\bibitem{mook} H. A. Mook, M. Yethiraj, G. Aeppli, T. E. Mason, and
               T. Armstrong
  Phys. Rev. Lett. {\bf 70}, 3490 (1993).
\bibitem{fong} H. F. Fong, B. Keimer, P. W. Anderson,
               D. Reznik, F. Dogan, and I. A. Aksay
  Phys. Rev. Lett. {\bf 75}, 316 (1995).
\bibitem{pwa} P.W. Anderson, Science, {\bf 235}, 1196 (1987).
\bibitem{rice} F.C. Zhang and T.M. Rice, 
  Phys. Rev. B {\bf 37}, 3759 (1988). 
\bibitem{demler} E. Demler and S.-C. Zhang,
  Phys. Rev. Lett. {\bf 75}, 4126 (1995).
\bibitem{zhang}S.-C. Zhang, Science, {\bf 275}, 1089 (1997).
\bibitem{baskaran}G. Baskaran and P.W. Anderson, 
  Report No. cond-mat/9706076.
\bibitem{greiter}M. Greiter, 
  Phys. Rev. Lett. {\bf 79}, 4898 (1997), and
  Report No. cond-mat/9705282.
\bibitem{zhang1} E. Demler, S.-C. Zhang, S. Meixner, and W, Hanke
  Phys. Rev. Lett. {\bf 79}, 4937 (1997)
\bibitem{laughlin}R.B. Laughlin, Report No. cond-mat/9709197.
\bibitem{arovas}D. Arovas, A. J. Berlinsky, C. Kallin and S.-C. Zhang,
  Phys. Rev. Lett. {\bf 79}, 2871 (1997).
\bibitem{andersen} N. H. Andersen, unpublished work at Ris\o\ National
Laboratory (1998).
\bibitem{dagotto} E. Dagotto, Rev. Mod. Phys. {\bf 66}, 763 (1994),
  and references therein. 
\bibitem{corr} D. N. Zheng, A. M. Cambell, J. D. Johnson,
  J. R. Cooper, F. J. Blunt, A. Porch, and P. A. Freeman,
  Phys. Rev. B {\bf 49}, 1417 (1994).
\bibitem{pumpin} B. P\"{u}mpin, H. Keller, W. K\"{u}ndig, W. Odermatt,
  I.M. Savi\'{c}, J.W. Schneider, H. Simmler, P. Zimmermann,
  E. Kaldis, S. Rusiecki, Y. Maeno, and C. Rossel,
  Phys. Rev. B {\bf 42}, 8019 (1990). 
\bibitem{chakravarty} S. Chakravarty, B. I. Halperin, and D. R. Nelson,
  Phys. Rev. B {\bf 39}, 2344 (1989). 
\bibitem{manousakis} E. Manousakis,
  Rev. Mod. Phys. {\bf 63}, 1 (1991),
\bibitem{igarashi} J. Igarashi,
  Phys. Rev. B {\bf 46}, 10 763 (1992). 
\bibitem {hayden91} S.M. Hayden, G. Aeppli, A.D. Taylor, T.G. Perring,
  S-W.  Cheong, and Z. Fisk, 
  Phys. Rev. Lett. {\bf 67}, 3622 (1991).
\bibitem{keimer} B.\ Keimer {\it et al.},
  Phys. Rev. B {\bf 46}, 14 034 (1992). 
\bibitem{hayden96}  S.M. Hayden, G. Aeppli, T.G. Perring,
  H. A. Mook, and F. Dogan,
  Phys. Rev. B {\bf 54}, R6905 (1996). 
\bibitem{pwa63} P. W. Anderson,
  Phys. Rev. {\bf 130}, 439 (1963).
\bibitem{gan} J. Gan and P. Hedeg\aa rd
  Phys. Rev. B {\bf 53}, 911 (1996). 
\bibitem{fradkin} E. Fradkin, 
  {\em Field Theories of Condensed Matter  Systems},
  Frontiers in Physics Vol.\ 82 
  (Addison-Wesley Publishing Company, New York, 1991).
\bibitem {cheong} S-W.  Cheong, Thompson, and Z. Fisk,
  Phys. Rev. B {\bf 39}, 4395 (1989). 
\end{references}
\end{document}